\documentclass[11pt] {article}
\usepackage{times}
\usepackage{graphics}
\usepackage{tensor}
\usepackage{natbib}
\usepackage{makeidx}
\usepackage{latexsym}
\usepackage{bm}
\usepackage{amsmath}
\usepackage{amsthm}
\usepackage{amssymb}
\makeindex

\newcommand{\beq}{\begin{equation}}
\newcommand{\eeq}{\end{equation}}

\newcommand{\defn}{\begin{quote}{\bf Definition. }}
\newcommand{\edefn}{\end{quote}}
\newcommand{\thm}{\begin{theorem}}
\newcommand{\ethm}{\end{theorem}}

\newcommand{\bmat}[1]{\left ( \begin{array}{#1}}
\newcommand{\emat}{\end{array}\right )}

\newcommand{\ts}{^{\sf T}} 
\newcommand{\its}{^{\sf -T}}

\newcommand{\bp}{{\bm \beta}}

\theoremstyle{definition}

\theoremstyle{plain}
\newtheorem{theorem}{Theorem}

\newcommand{\eps}[3]
{{\begin{center}
 \rotatebox{#1}{\scalebox{#2}{\includegraphics{#3}}}
 \end{center}}
}

\newcommand{\dsp}{1}

\renewcommand{\baselinestretch}{\dsp}

\begin{document}

\title{P-splines with derivative based penalties and tensor product smoothing of unevenly distributed data.}
\author{ Simon N. Wood\\ School of Mathematics, University of Bristol, Bristol, U.K.\\
{\tt simon.wood@bath.edu}}

\maketitle

\renewcommand{\baselinestretch}{1}

\begin{abstract}
The P-splines of \cite{Eilers&Marx96} combine a B-spline basis with a discrete quadratic penalty on the basis coefficients, to produce a reduced rank spline like smoother. P-splines have three properties that make them very popular as reduced rank smoothers: i) the basis and the penalty are sparse, enabling efficient computation, especially for Bayesian stochastic simulation; ii) it is possible to flexibly `mix-and-match' the order of B-spline basis and penalty, rather than the order of penalty controlling the order of the basis as in spline smoothing; iii) it is very easy to set up the B-spline basis functions and penalties. The discrete penalties are somewhat less interpretable in terms of function shape than the traditional derivative based spline penalties, but tend towards penalties proportional to traditional spline penalties in the limit of large basis size. However part of the point of P-splines is not to use a large basis size. In addition the spline basis functions arise from solving functional optimization problems involving derivative based penalties, so moving to discrete penalties for smoothing may not always be desirable. The purpose of this note is to point out that the three properties of basis-penalty sparsity, mix-and-match penalization and ease of setup are readily obtainable with B-splines subject to derivative based penalization. The penalty setup typically requires a few lines of code, rather than the two lines typically required for P-splines, but this one off disadvantage seems to be the only one associated with using derivative based penalties. As an example application, it is shown how basis-penalty sparsity enables efficient computation with tensor product smoothers of scattered data.
\end{abstract}

\renewcommand{\baselinestretch}{\dsp}

\section{Computing arbitrary derivative penalties for B-splines}

The main purpose of this note is to show that reduced rank spline smoothers with derivative based penalties can be set up almost as easily as the P-splines of \cite{Eilers&Marx96}, while retaining sparsity of the basis and penalty and the ability to mix-and-match the orders of spline basis functions and penalties. The key idea is that we want to represent a smooth function $f(x)$ using a rank $k$ spline basis expansion  $f(x) = \sum_{j=1}^k \beta_j B_{m_1,j}(x)$, where $B_{m_1,j}(x)$ is an order $m_1$ B-spline basis function, and $\beta_j $ is a coefficient to be estimated. In this paper order $m_1 = 3$ will denote a cubic spline. Associated with the spline will be a derivative based penalty
$$
J = \int_a^b f^{[m_2]}(x)^2 dx 
$$
where $f^{[m_2]}(x)$ denotes the $m_2^{\text{th}}$ derivative of $f$ with respect to $x$, and $[a,b]$ is the interval over which the spline is to be evaluated. It is assumed that $m_2 \le m_1$, otherwise the penalty is formulated in terms of a derivative that is not properly defined for the basis functions, which makes no sense. It is possible to write $J=\bp \ts {\bf S}\bp$ where $\bf S$ is a band diagonal matrix of known coefficients. Computation of $\bf S$ is the only part of setting up the smoother that presents any difficulty, since standard routines for evaluating B-splines basis functions (and their derivatives) are readily and widely available, and in any case the recursion for basis function evaluation is straightforward.

The algorithm for finding $\bf S$ in general is as follows. $ p = m_1-m_2 $ denotes the order of piecewise polynomial defining the $m_2^{\text{th}}$ derivative of the spline. Let $x_1, x_2 \ldots x_{k-m+1}$ be the (ordered) `interior knots' defining the B-spline basis, that is the knots within whose range the spline and its penalty are to be evaluated (so $a=x_1$ and $b=x_{k-m+1}$). Let the inter-knot distances be $h_j = x_{j+1}-x_j$, for $0<j\le k-m$.
\begin{enumerate}
\item For each interval $[x_{j},x_{j+1}]$, generate $p+1$ evenly spaced points within the interval. For $p=0$ the point should be at the interval centre, otherwise the points always include the end points $x_j$ and $x_{j+1}$. Let ${\bf x}^\prime$ contain the unique $x$ values so generated, in ascending order. 
\item Obtain the matrix $\bf G$ mapping the spline coefficients to the $m_2^\text{th}$ derivative of the spline at the points ${\bf x}^\prime$.
\item If $p=0$, ${\bf W} = \text{diag}({\bf h})$.
\item It $p>0$, let $p+1 \times p+1$ matrices ${\bf P}$ and $\bf H$ have elements $P_{ij} = (-1 + 2(i-1)/p)^j$ and $H_{ij} =  (1+(-1)^{i+j-2})/(i+j-1)$ ($i$ and $j$ start at 1). Then compute matrix $\tilde {\bf W} = {\bf P}\its {\bf HP}^{-1}$. Now compute ${\bf W} = \sum_{q} {\bf W}^q$ where each ${\bf W}^q$ is zero everywhere except at $W^q_{i + pq - p, j + pq - p} = h_q \tilde W_{ij}/2$, for $i=1,\ldots,p+1$, $j=1,\ldots,p+1$. $\bf W$ is banded with $2p + 1$ non-zero diagonals.
\item The diagonally banded penalty coefficient matrix is ${\bf S} = {\bf G}\ts {\bf WG}$. 
\item Optionally, compute the diagonally banded Cholesky decomposition ${\bf R}\ts {\bf R} = {\bf W}$, and form diagonally banded matrix ${\bf D} = {\bf RG}$, such that ${\bf S} = {\bf D}\ts {\bf D}$.  
\end{enumerate}
Step 2 can be accomplished by standard routines for generating B-spline bases and their derivatives of arbitrary order: in {\tt R} for example, the function {\tt splines:splineDesign} 
for normal B-splines or {\tt mgcv:cSplineDes} for cyclic B-splines. Alternatively see the appendix. Step 4 requires no more than a single rank $p+1$ matrix inversion of $\bf P$. ${\bf P}$ is somewhat ill conditioned for $p \ge 20$, with breakdown for $p > 30$. However it is difficult to imagine any sane application for which $p$ would even be as high as 10, and for $p\le 10$, $\bf P$'s condition number is $< 2 \times 10^4$. Of course $\bf W$ is formed without explicitly forming the ${\bf W}^q$ matrices. Step 6 can be accomplished by a banded Cholesky decomposition such as {\tt dpbtrf} from LAPACK (accessible via routine {\tt mgcv:bandchol} in {\tt R}, for example). Alternatively see the appendix. However for applications with $k$ less than 1000 or so, a dense Cholesky decomposition might be deemed efficient enough. Note that step 6 is preferable to construction of $\bf D$ by decomposition of $\bf S$, since $\bf W$ is positive definite by construction, while, for $m_2>0$, $\bf S$ is only positive semi-definite. As in the case of a discrete P-spline penalty the leading order computational cost of evaluating $\bf S$ is $O(bk)$ where b is the number of bands in $\bf S$ (the $O(p^3)$ cost of $\tilde {\bf W}$ usually being negligible in comparison), and is a trivial relative to model fitting. 

The derivation of the algorithm is quite straightforward. Given the basis expansion we have that
$$
S_{ij} = \int_a^b B_{m_1,i}^{[m_2]}(x)B_{m_1,j}^{[m_2]}(x) dx.
$$
However by construction $B_{m_1,i}^{[m_2]}(x)$ is made up of order $p = m_1-m_2$ polynomial segments. So we are really interested in integrals of the form
$$
S_{ijl} = \int_{x_l}^{x_{l+1}} B_{m_1,i}^{[m_2]}(x)B_{m_1,j}^{[m_2]}(x) dx = 
\frac{h_l}{2}\int^1_{-1}  \sum_{i=0}^p a_i x^i \sum_{j=0}^p d_j x^j dx
$$
for some polynomial coefficients $a_i$ and $d_j$. The polynomial coefficients are the solution obtained by evaluating $B_{m_1,i}^{[m_2]}(x)$ at $p+1$ points spaced evenly from $x_l$ to $x_{l+1}$, to obtain a vector of evaluated derivatives, ${\bf g}_a$, and then solving ${\bf Pa} = {\bf g}_a$ ($\bf d$ is obtained from ${\bf g}_d$ similarly). Then $S_{ij} = \sum_l S_{ijl}$.

Given that $\int_{-1}^1 x^q dx = (1+ (-1)^q)/(q+1) $ it is clear that $S_{ijl} = h_l{\bf a}\ts {\bf H} {\bf d}/2$ where $H_{ij} = (1+(-1)^{i+j-2})/(i+j-1)$ ($i$ and $j$ start at 1). In terms of the evaluated gradient vectors, 
$$
S_{ijl} = h_l {\bf g}_a\ts {\bf P}\its {\bf HP}^{-1}  {\bf g}_d/2.
$$
The $\bf G$ matrix simply maps $\bp$ to the concatenated (and duplicate deleted) gradient vectors for all intervals, while $\bf W$ is just the overlapping-block diagonal matrix with blocks given by $h_l  {\bf P}\its {\bf HP}^{-1}/2$, hence $S_{ij} = {\bf G}_i\ts {\bf WG}_j$, where ${\bf G}_i$ is the $i^\text{th}$ row of $\bf G$. The simplicity of the algorithm rests on the ease with which $\bf G$ and $\bf W$ can be computed. Note that the construction is more general than that of \cite{wandormerod2008}, in allowing $m_1$ and $m_2$ to be chosen freely (rather than $m_1$ determining $m_2$), and treating even $m_1$ as well as odd.

\section{Tensor product smoothing of unevenly distributed data \label{te}}

An example where a compactly supported basis and sparse penalty is computationally helpful is  in tensor product smoothing of unevenly distributed data. A three dimensional example suffices to illustrate how tensor product smooths are constructed from one dimensional bases. 
Suppose we want to smooth with respect to $z_1$, $z_2$ and $z_3$. Firstly B-spline bases are constructed for smooth functions of each covariate separately. Suppressing subscripts for order, let $B_{j1}(z_j), B_{j2}(z_j), \ldots$ denote the basis for the smooth function of $z_j$, and let ${\bf D}_j$ denote the corresponding `square root' penalty matrix. The smooth function of all three variables is then represented as
$$
f({\bf z}) = \sum_{ijl} \beta_{ijl} B_{1i}(z_1)B_{2j}(z_2)B_{3l}(z_3)
$$   
where $\beta_{ijl}$ are the coefficients. Notice that the tensor product basis functions, $B_{1i}(z_1)B_{2j}(z_2)B_{3l}(z_3)$, inherit compact support form the marginal basis functions. Now write the coefficients in `column major' order in one vector $\bp\ts = (\beta_{111},\beta_{112},\ldots, \beta_{11k_1}, \beta_{121}, \beta_{122},\ldots \beta_{k_1k_2k_3})$, where $k_j$ is the dimension of the $j^{\rm th}$ basis. The tensor product smoother then has three associated penalties, $\bp \ts {\bf S}_j\bp$ (each with its own smoothing parameter), where ${\bf S}_j = \tilde {\bf D}_j\ts \tilde {\bf D}_j$,
$$
\tilde {\bf D}_1 =  {\bf D}_1 \otimes {\bf I}_{k_2} \otimes {\bf I}_{k_3}, ~~
\tilde {\bf D}_2 = {\bf I}_{k_1} \otimes {\bf D}_2 \otimes  {\bf I}_{k_3} \text{ and }
\tilde {\bf D}_3 = {\bf I}_{k_1}\otimes  {\bf I}_{k_2} \otimes {\bf D}_3 .
$$
This construction generalizes to other numbers of dimensions in the obvious way \citep[see e.g.][]{wood2006tensor}. 

By construction the domain of the tensor product smooth is a rectangle, cuboid or hypercuboid, but it is often the case that the covariates to be smoothed over occupy only part of this domain. In this case it is possible for some basis functions to evaluate to zero at every covariate observation, and there is often little point in retaining these basis functions and their associated coefficients. Let $\iota$ denote the index of a coefficient to be dropped from $\bp$ (along with its corresponding basis function). The n\"aive approach of dropping row and column $\iota$ of each ${\bf S}_j$ is equivalent to setting $\beta_\iota$ to zero when evaluating $\bp \ts {\bf S}_j\bp$, which is not usually desirable. Rather than setting $\beta_\iota=0$ in the penalty, we would like to omit those components of the penalty dependent on $\beta_\iota$. This is easily achieved by dropping every row $\kappa $ from $\tilde {\bf D}_j$ for which $\tilde D_{j,\kappa \iota} \ne 0$. Notice (i) that without $\bf D$ being diagonally banded this would be a rather drastic reduction of the penalty, and (ii) this construction applies equally well to P-splines.

\renewcommand{\baselinestretch}{1}

\begin{figure}
\vspace*{-1cm}
\eps{-90}{.6}{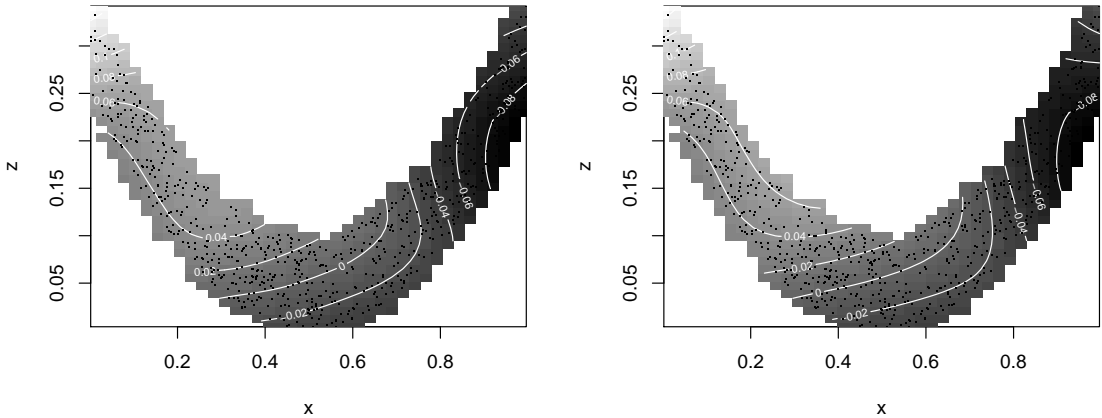}
\vspace*{-.5cm}
\caption{Left: conventional tensor product smooth reconstruction of the example function given in the text, based on noisy samples at the $x, z$ locations shown as black dots. Right: as left, but using the reduced basis described in section \ref{te}.
 \label{bs.fig}}
\end{figure}
\renewcommand{\baselinestretch}{\dsp}

As an illustration data were generated from the model
$$
y_i = \exp\{-(z_i-0.3)^2/2 - (x_i-0.2)^2/4\} + \epsilon_i, \text{ where } \epsilon_i \sim N(0,0.1^2)
$$
at the $x, z$ locations shown as black dots in figure \ref{bs.fig}. The figure shows the reconstruction of the test function using a tensor product smoother, based on cubic spline marginals with second derivative penalties. The left figure is for the full smoother, which had 625 coefficients, while the right figure is for the reduced version which had 358 coefficients. Including REML smoothing parameter selection the reduced rank fit took around 1/8 of the computation time of the full rank fit. The correlation between the fitted values for the two fits is 0.999. In the example the reduced rank fit has marginally smaller mean square reconstruction error than the full rank version, a feature that seems to be robust under repeated replication of the experiment. 
 
\section{Conclusions}

Given that the theoretical justification for using spline bases for smoothing is that they arise as the solutions to variational problems with derivative based penalties \citep[see e.g.][]{wahba90, duchon77}, it is sometimes appealing to be able to use derivative based penalties for reduced rank smoothing also. However if a sparse smoothing basis and penalty were required alongside the ability to mix-and-match penalty order and basis order, then the apparent complexity of obtaining the penalty matrix for derivative based penalties has hitherto presented an obstacle to their use. This note removes this obstacle, allowing the statistician an essentially free choice whether to use derivative based penalties or discrete penalties. The splines described here are available in {\tt R} package {\tt mgcv} from version 1.8-12. They could be referred to as `D-splines', but a new name is probably un-necessary. This work was supported by EPSRC grant EP/K005251/1.

\appendix

\section{Standard recursions}

B-spline bases, their derivatives and banded Cholesky decompositions are readily available in standard software libraries and packages such as {\tt R} and {\tt Matlab}. However, for completeness the required recursions are included here. 

To define a $k$ dimensional B-spline basis of order $m$ we need to define $k+m+1$ knots $x_1 < x_2 < \ldots < x_{k+m+1}$. The interval over which the spline is to be evaluated is $[x_{m+1},x_{k+1}]$ so the locations of knots outside this interval are rather unimportant. The B-spline basis functions are defined recursively as
$$
B_{m,i}(x) = \frac{x-x_i}{x_{i+m} - x_i} B_{m-1,i}(x) + 
\frac{x_{i+m+1}-x}{x_{i+m+1} - x_{i+1}} B_{m-1,i+1}(x), ~~~ i=1,\ldots, k, ~~ m > 0
$$ 
where 
$$
B_{0,i}(x) = \left \{ \begin{array}{ll} 
1 & x_i \le x < x_{i+1} \\ 0 & \text{otherwise}.
\end{array}
\right .
$$
It turns out that the derivative with respect to $x$ of a B-spline of order $m$ can be expressed in terms of a B-spline basis of order $m-1$ as follows
$$
\sum_j \beta_j B_{m,j}^\prime(x) = (m-1) \sum_j \frac{\beta_j - \beta_{j-1}}{x_{j + m} - x_j} 
B_{m-1,j}(x).
$$
This can be applied recursively to obtain higher order derivatives. For more on both of these recursions see and p.89 and p.116 of \cite{deBoor2001} \citep[or][]{deboor78}.

Now consider the banded Cholesky decomposition of a symmetric positive definite matrix $\bf A$ with $2p-1$ non-zero diagonals (clustered around the leading diagonal). We have
$$
R_{ii} = \sqrt{A_{ii} - \sum_{k=i-p}^{i-1} R_{ki}^2 }, \text{~and~}
R_{ij} = \frac{A_{ij} - \sum_{k=i-p}^{i-1} R_{ki}R_{kj}}{R_{ii}}, ~~~ i < j < i + p.
$$ 
all other elements of Cholesky factor ${\bf R}$ being 0. The expressions are used one row at a time, starting from row 1, and working across the columns from left to right. See any matrix algebra book for Cholesky decomposition \citep[e.g.][]{golub.vanloan}.

\bibliography{/home/sw283/bibliography/journal,/home/sw283/bibliography/simon}
\bibliographystyle{chicago}

\end{document}